\documentclass[letterpaper, 10 pt, conference]{ieeeconf}  

\IEEEoverridecommandlockouts                              

\usepackage{graphics} 
\usepackage{epsfig} 
\usepackage{amsmath} 
\usepackage{amssymb}  
\usepackage{caption}

\usepackage{verbatim}
\usepackage{graphicx} 
\usepackage{epsfig} 
\usepackage{epstopdf}
\usepackage{amsmath} 
\usepackage{amssymb}  
\usepackage{color}

\makeatletter
\newcommand{\figcaption}{\def\@captype{figure}\caption}
\newcommand{\tabcaption}{\def\@captype{table}\caption}

\usepackage{hyperref}

\makeatother
\title{\LARGE \bf   Rebalancing Frequency Considerations  for  Kelly-Optimal\\ Stock Portfolios in a Control-Theoretic Framework
}

\author{\large Chung-Han Hsieh,$^{1}$  John A. Gubner,$^{2}$ and B. Ross Barmish$^{3}$
	\thanks{\hskip -10pt ${}^1$Chung-Han Hsieh is a graduate student working towards to his Ph.D. degree in the Department of Electrical and Computer Engineering, University of Wisconsin, Madison, WI 53706. E-mail: hsieh23@wisc.edu.}
	\thanks{\hskip -10pt ${}^2$John A. Gubner is a faculty member in  the Department of Electrical and Computer Engineering, University of Wisconsin, Madison, WI 53706. \mbox{E-mail}: john.gubner@wisc.edu.}
	\thanks{\hskip -10pt ${}^3$B. Ross Barmish is a faculty member in  the Department of Electrical and Computer Engineering, University of Wisconsin, Madison, WI 53706. \mbox{E-mail}: barmish@engr.wisc.edu.}
}

\begin{document}

	\maketitle
	\thispagestyle{empty}
	\pagestyle{empty}
	
	\parindent = 0pt
	
	\begin{abstract}
		In this paper, motivated by the celebrated work of Kelly, we consider the problem of portfolio weight selection to
		maximize expected logarithmic growth. Going beyond existing literature, our focal point here is the {\itshape rebalancing frequency\/}  which we include as an additional parameter in our analysis.
		The problem is first set in a control-theoretic framework, and then, the main question we address is as follows: In the absence of transaction costs, does high-frequency trading always lead to the best performance?
		Related to this is our prior work on betting, also in the Kelly context, which examines the impact of making a wager and letting it ride. Our results on betting frequency   can be interpreted in the context of weight selection for a two-asset portfolio consisting of one risky asset and one riskless asset.
		With regard to the question above, our prior results indicate that it is often the case that there are no performance benefits associated with high-frequency trading.
		In the present paper, we generalize the analysis to   portfolios with multiple risky assets. We show that if there is an asset satisfying
		a new condition which we call {\itshape dominance}, then an optimal portfolio consists of this asset alone; i.e., the trader has ``all eggs in one basket'' and  performance becomes a constant function of rebalancing frequency. Said another way, the problem of rebalancing is rendered moot. The paper also includes simulations which address
		practical considerations associated with real stock prices
		and the dominant asset condition.
	\end{abstract}


	\vspace{6mm}
	\section{Introduction}
	\label{SECTION: INTRODUCTION}
	\vspace{-1mm}
	
	The main results of this paper pertain to the effect of ``rebalancing frequency'' for portfolio weight selection problems with performance measured using Kelly's celebrated expected logarithmic growth criterion, which  was first used for  a variety of sequential betting problems~\cite{Kelly_1956}; see also~\mbox{\cite{Latane_1959}-\cite{Cover_Thomas_2012}} where further results along these lines are given. In this regard, the work reported in this paper is part of a line of research using this criterion in the context  of portfolio optimization in the stock market;~e.g., see~\cite{Luenberger_2011} and \cite{Thorp_2006} for a good introduction,  \cite{Maclean_Thorp_Ziemba_2010} for a rather comprehensive exposition on the properties of solutions obtained using expected logarithmic growth, and~\mbox{\cite{Kuhn_Luenberger_2010}-\cite{MacLean_Thorp_Ziemba_2011}} for a sampling of some more recent developments. Initial results about rebalancing frequency are reported in~\cite{Kuhn_Luenberger_2010} and~\cite{Das_Kaznachey_Goyal_2014} for the case when the stock prices follow a continuous-time geometric Brownian motion. Additionally, a drawback in~\cite{Das_Kaznachey_Goyal_2014} is that the betting fraction $K$ is chosen without regard to the frequency at which the portfolio is rebalanced. Subsequently, when this same fraction $K$ is used to find an optimal rebalancing period, the resulting levels of logarithmic growth are suboptimal. To complete this overview,
	we single  out~\cite{MacLean_Thorp_Ziemba_2011} for emphasis since it provides a comprehensive survey  covering many of the most important papers in this line of research. 
	
	\vspace{4mm} 
	Most closely related to this paper is our recent work~\cite{Hsieh_Barmish_Gubner_2018_ACC} which considers a repeated betting game and the impact on expected logarithmic growth resulting from making a wager and letting it
	ride for several steps in lieu of updating. This can be interpreted as
	weight selection for a two-asset portfolio and  ``letting it ride" to capture  the
	effect of the frequency of rebalancing. With this as  background, this paper is aimed at generalizing these initial results on frequency dependence   to trading a multi-stock portfolio.

	\vspace{4mm}
	The appeal of this research to the control community is based on the fact that the Kelly-based  rebalancing problem can be formulated as a stochastic control problem with a linear feedback  and  randomly varying  inputs corresponding to the vector of stagewise returns~$X(k)$ on  portfolio assets; see~\mbox{\cite{Hsieh_Barmisg_2015_Allerton}-\cite{Hsieh_Barmish_Gubner_2016_CDC},  \cite{Calafiore_Monastero_2010}-\cite{Hsieh_Barmish_2017_CDC}},
	and~\cite{Barmish_Primbs_TAC_2016} where a similar control-theoretic set-up is considered  for finance problems in continuous time. To study the effect of rebalancing  frequency for portfolio problems, let~$\Delta t$ be the time between portfolio updates. With~$n$ being the number of steps between rebalancings,   the  {\it  frequency} is 
	$$
	f \doteq \frac{1}{ n \Delta t}.
	$$  Subsequently,  
	for each $n$, the expected logarithmic growth using optimal
	portfolio weights is denoted by~$g_n^*$, which we study as a function of $n$.
	
	\vspace{4mm}
	The main questions we address in this paper are as follows:  Does   high-frequency trading, corresponding to~$n = 1$, always lead to the best performance? Under what conditions can a low-frequency trader using~\mbox{$n > 1$} match or exceed the optimal high-frequency performance level~$g_1^*$? Indeed, in the presence of transaction costs, our previous work mentioned above, carried out in the context of sequential betting, includes  a demonstration that~\mbox{$g_n^* > g_1^*$} is possible when transaction costs are in play. That is, the prohibitive costs associated with trading too often  may render   high-frequency trading suboptimal. However, for the zero transaction-cost case, we also showed  that it is possible to obtain~\mbox{$g_1^* = g_n^*$} for all $n \geq 1$ although  it is still an open question whether~\mbox{$g_n^* > g_1^*$} is possible. For this case, in the sequel, we generalize these results in~\cite{Hsieh_Barmish_Gubner_2018_ACC} to the multiple-risky-asset case, and prove that  there are many scenarios where the low-frequency trader's performance can actually match that of the high-frequency trader~--- the extreme case with~$n$ very large corresponding to buy and hold. This performance matching is proven  when at least one of the assets in the portfolio is {\it dominant\/} in the sense that it is relatively more attractive than every other potential asset under consideration. In this case,
	it becomes arguable that dynamic portfolio rebalancing is a ``waste of~time'' to even consider.

	\vspace{4mm}
	To complete this overview, we should also mention  another result in the literature  involving rebalancing frequency considerations. In~\cite{Kuhn_Luenberger_2010},  the returns are assumed to follow  a continuous-time geometric Brownian motion  and two extreme cases are considered~--- when the time between rebalancing is either very large or very small.  
	In contrast to~\cite{Kuhn_Luenberger_2010}, we consider the entire range of frequencies from low to high. To this end, our objective here is to analyze, in discrete time, the more general case when both the probability distribution of the returns and the time interval between updates are~arbitrary. 
	
	\vspace{6mm}
	{\bf Preview of Main Result:} 
	For the case when the portfolio is comprised of two or more potentially investable assets with each having i.i.d. returns $X_i(k)$ and possibly correlated, Asset~$j$ is said to be {\it dominant}~if 
	$$
	\mathbb{E}\left[ {\frac{1+X_i(0)}{{1 + X_j(0)}}} \right] \leq 1
	$$
	holds for all~$i \neq j$. In this case, our main result, which we call the {\it Dominant Asset Theorem}, tells us that when this condition is satisfied, an optimal strategy is obtained by investing all of the trader's funds in Asset~$j$.  Figuratively speaking, this result says that an optimal portfolio is obtained by putting all  eggs in one basket.  Of equal importance, as a consequence of the theorem, it is seen that  the performance of the high-frequency trader and the buy and holder are  identical. That is,~$g_n^* = g_1^*$ for all~$n \geq 1$. Thus, performance is invariant to the rebalancing frequency and it follows that there is no benefit associated with trading often;  it suffices to buy~and~hold. Said another way, if all funds are invested in a single asset, which could be cash, then rebalancing is rendered moot. Equivalently,  the  performance must be a constant function of $n$.
	
	\vspace{5mm}
	{\bf Theoretical Versus Practical Considerations}: 
	Consistent with the vast preponderance of results in the literature, our approach is model based in the sense that the probability distribution for the asset returns is known; see Section~\ref{SECTION: PROBLEM FORMULATION} for details.
	The reader is also referred to~\cite{Luenberger_2011} and \cite{Cox_Ross_Rubinstein_1979} where binomial lattice models are used to approximate geometric Brownian motion. 
	In practice, the probability distribution of the returns is typically  estimated from historical data. In view of the fact that real-world stock returns are generally nonstationary, in practice, frequent updates of the model accompanied by portfolio rebalancing are in order. That is, at best, model-based results should  be viewed as useful only for a limited amount of time.   In Section~\ref{SECTION: EMPIRICAL DATA STUDY}, the reader is provided with the flavor  of these practical issues in the context of the dominance result described above.

	\vspace{6mm}
	\section{Control-Theoretic Formulation}
	\label{SECTION: PROBLEM FORMULATION}
	\vspace{-0mm}
	In this section, we begin with a trader who is forming a portfolio and considering $m \geq 2$ potential assets for inclusion. We now formulate  the frequency-dependent portfolio problem in control-theoretic terms.  Indeed, the system output at stage $k$ is taken to be the trader's time-varying account value~$V(k)$ for~\mbox{$k=0,1,\ldots,n-1,$} and, for $i=1,2,\ldots, m$, we use feedback gains~\mbox{$0 \leq K_i \leq 1$} to represent the fraction of this amount  allocated to the~\mbox{$i$-th} asset. The inequality~$K_i \geq 0$ means that the trader is going {\it long} and   {\it short selling} is disallowed. In addition,~\mbox{$K_i \leq 1$} forces the amount invested  to be no more than the account value~$V(k)$. In other words, this disallows the use of {\it leverage} and possible margin costs. This no-leverage requirement, applied to the portfolio in its entirety, leads to the constraint
	$$
	K \in {\cal K} \doteq \left\{K \in \mathbb{R}^{m}: K_i \geq 0, \; \sum_{i=1}^m K_i = 1 \right\}.
	$$ 
	That is, with~\mbox{$K \in \cal K$}, we have a guarantee that no more than~100\% of the account value $V(k)$ is invested; see the conclusion for further discussion.  Now, the~$i$-th control signal  is  a linear feedback of the form  
	\[
	I_i(k) = K_iV(k)
	\]
	which is called the $i$-th {\it investment function.}
	
	\vspace{5mm}
	{\bf The Asset Returns}:
	If Asset~$i$ is a stock whose price at time~$k$ is $S_i(k)$, then its return is
	\[
	X_i(k) \doteq \frac{S_i(k+1) - S_i(k)}{S_i(k)}.
	\]
	In the sequel, we assume a ``perfect model" for the stochastic process driving the
	stock prices. That is, for  risky assets, we assume that the return vectors $X(k)$ have a known distribution with components $X_i(k)$ which can be arbitrarily correlated. It is also assumed that these vectors are  independent and identically distributed~(i.i.d.) with components satisfying
	$$
	X_{\min,i}  \leq X_i(k) \leq X_{\max,i}
	$$
	with known bounds above and with $X_{\max,i}$ being finite and ~\mbox{$X_{\min,i} > -1$}. This means that the loss per time step is limited to less than~$100\%$ and is interpreted to mean that the price of a stock cannot drop to zero. To avoid triviality, we assume that at least one of the $m$ assets is riskless with nonnegative rate of return $r_i \geq 0$. That is, if Asset $i$ is riskless, its return is deterministic and is treated as a degenerate random variable~\mbox{$X_i(k) = r_i  $} for all $k$ with probability one.  The quantity $r_i$ is called an  interest rate, and it is noted that this formulation also allows for the trader to maintain cash in the account by taking $r_i=0.$\footnote{ There are  rare cases when the best possible riskless asset has negative returns and the optimal portfolio, which will be discussed in Section~\ref{SECTION: DOMINATE ASSET THEOREM}, might be one which involves losing money as slowly as possible.  }

	\begin{figure}[htbp]
		\vspace{-0mm}
		\begin{center}
			\graphicspath{{Figs/}}
			\includegraphics[scale=0.33]{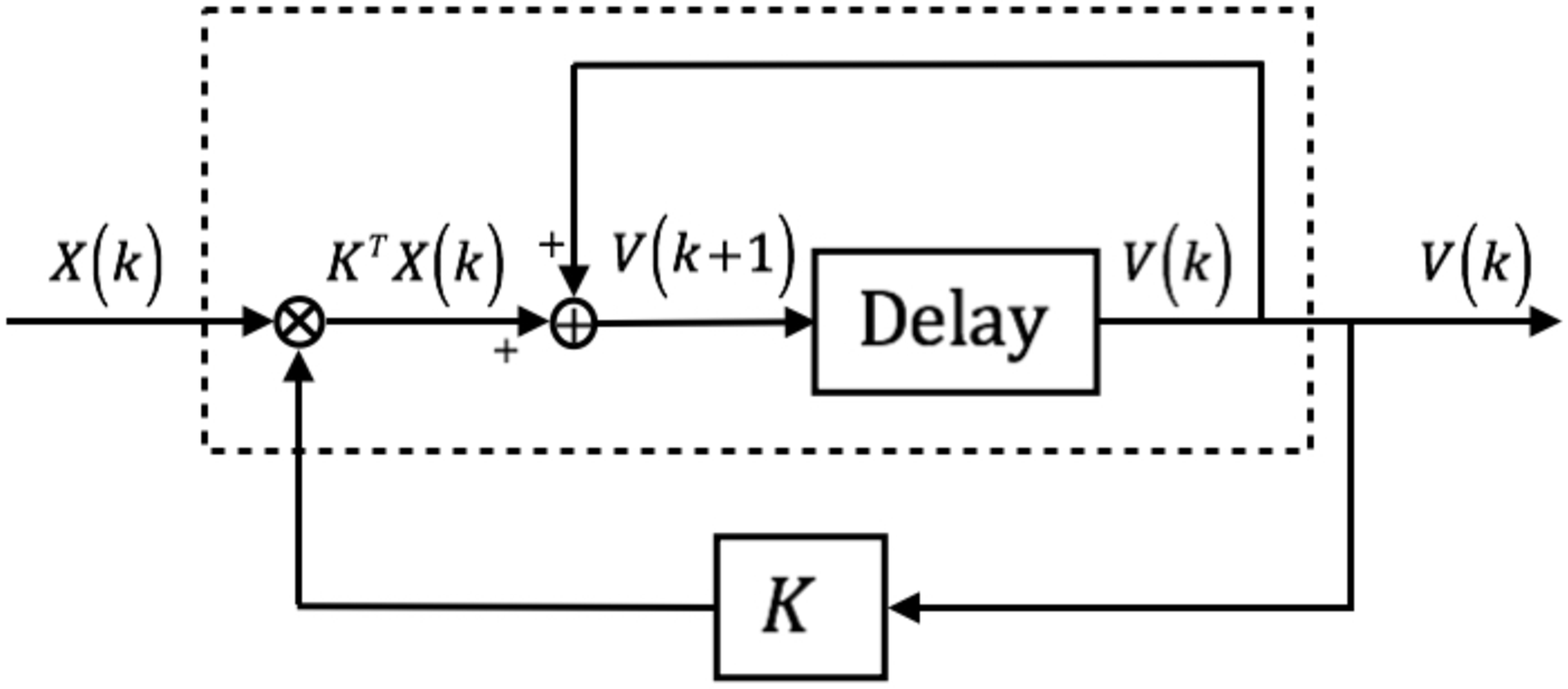}
			\figcaption{Feedback Configuration for Trading}
			\label{fig:Block_Diagram_KV}
		\end{center}
		\vspace{-0mm}
	\end{figure}

	\vspace{3mm}
	{\bf Dynamics and Trading Frequency Considerations}: 
	The update in account value from stage $k$ to $k+1$ for the resulting closed-loop system, depicted in Figure~\ref{fig:Block_Diagram_KV}, is
	\begin{align*}
	V(k+1)&= (1+K^TX(k))V(k).   
	\end{align*}
	Letting $n$ be the number of steps between rebalancings,  at time~\mbox{$k=0$}, the trader begins with investment~\mbox{$I(0) = KV(0)$}
	and waits~$n$ steps in the spirit of ``buy and hold."  Then, when~\mbox{$k=n$}, the investment is updated to be
	$$
	I(n) = KV(n).
	$$ 
	Continuing in this manner, a waiting period of $n$ stages is enforced between each rebalance. Now, to study performance as a function of frequency, we use the {\it compound~returns}
	\[
	\mathcal{X}_{n,i} \doteq \prod_{k=0}^{n-1} (1+X_i(k)) -1
	\]
	which are readily seen to satisfy~\mbox{$\mathcal{X}_{n,i} > -1$} for all $n$ and $i=1,2,\ldots,m$.
	In the sequel, we work with the random vector $\mathcal{X}_n$ having $i$-th
	component~$\mathcal{X}_{n,i}$.
	Then, for any fixed rebalancing period~$n$ and initial account value~$V(0)>0$, the corresponding account value at stage $n$ is given by
	\begin{align*}
	V(n) &\doteq (1 + K^T \mathcal{X}_n )V(0).  
	\end{align*}

	\vspace{5mm}
	{\bf The Frequency Dependent Optimization Problem}: As a function of $n \geq 1$, we 
	study the problem of maximizing   the expected logarithmic growth
	\begin{eqnarray*}
		{g_n}(K)\!\!\! &\doteq& \!\!\! \frac{1}{n}\mathbb{E}\left[ \log \left(\frac{V(n)}{V(0)}\right) \right] \\[5pt]
		&=&\!\!\! \frac{1}{n}\mathbb{E}\left[ {\log (1 + {K^T}{\mathcal{X}_n})} \right],
	\end{eqnarray*}
	which is concave in $K$. The associated optimal expected logarithmic growth is  obtained as
	\[
	g_n^* \doteq \max_{K \in \mathcal{K}}g_n(K)
	\]
	and any $K_n^* \in \mathcal{K}$ satisfying $g_n(K_n^*) = g_n^*$ is called an \mbox{\it optimal Kelly fraction} for the rebalancing period of length~$n$. 

	\vspace{6mm}
	\section{Relative Attractiveness and Dominance}
	\vspace{-0mm}
	
	As discussed in Section~\ref{SECTION: INTRODUCTION}, in our
	prior work~\cite{Hsieh_Barmish_Gubner_2018_ACC},  the Kelly betting problem  results    can be interpreted in the context of  weight selection for a two-asset portfolio consisting of one risky asset and one riskless asset.  We proved that under
	a simple condition which we called ``sufficient attractiveness,"~$g_n^*$ is a constant function of~$n$. Thus, when this condition holds, trading
	faster does~{\it not\/}   lead to performance improvement over a simple buy-and-hold strategy. To extend these results to the case of a portfolio of arbitrary size~$m$, we   generalize the notion of sufficient attractiveness with the definition below.

	\vspace{5mm}
	{\bf Definition (Relative Attractiveness and Dominance)}: Given a collection of $m$ assets, we say that Asset~$j$ is {\it relatively more attractive} than Asset~$i$ if 
	$$
	\mathbb{E}\left[ {\frac{1+X_i(0)}{{1 + X_j(0)}}} \right] \leq 1.
	$$
	Equivalently, Asset~$j$ is relatively more attractive than Asset~$i$ if the correlation between 
	$[1+X_j(0)]^{-1}$ and $1+X_i(0)$ is at most one.
	Asset~$j$ is said to be {\it dominant} if it is relatively more attractive
	than every other asset $i\ne j$. 
	
	\vspace{5mm}
	{\bf Remarks}: $(i)$
	When $m=2$, we note that Asset~$j$ is dominant if and only if it is
	relatively more attractive than Asset~$i$. $(ii)$
	If $m=2$ and Asset~$i$ is riskless with $X_i(0)=0$,
	then the dominance of Asset~$j$ is equivalent
	to its sufficient attractiveness as defined
	in~\cite{Hsieh_Barmish_Gubner_2018_ACC}. $(iii)$ If $m \geq 2$, a riskless Asset~$j$ with interest rate $r$ is easily seen to
	be relatively more attractive than
	risky Asset~$i$ if and only~if
	$$
	\mathbb{E}[X_i(0)] \le r. 
	$$
	$(iv)$ For a risky Asset~$j$ to be relatively more attractive
	than the riskless Asset~$i$, we require more than just~\mbox{$\mathbb{E}[X_j(0)]>r$.}
	For example, consider returns {$X_j(k) \in \{-1/2,1/2\}$}
	with 
	$$
	P(X_j(k)=1/2)=0.6.
	$$
	Then with $X_i(k)=r=0.05$,
	a straightforward calculation leads to
	$$\mathbb{E}[X_j(0)]>r,$$ but
	$$
	\mathbb{E}\left[ {\frac{1+X_i(0)}{{1 + X_j(0)}}} \right] = 1.26
	$$
	which violates the relative attractiveness inequality. $(v)$
	Although the condition $\mathbb{E}[X_j(0)]\geq r$ is not  sufficient
	for a risky Asset~$j$ to be relatively more attractive than
	a riskless Asset~$i$, the condition is necessary.
	This can be seen by
	applying Jensen's inequality to obtain
	\[\frac{{1 + r}}{{\mathbb{E}[1 + {X_j}(0)]}} \leq \mathbb{E}\left[ {\frac{{1 + r}}{{1 + {X_j}(0)}}} \right]. \] 
	If Asset~$j$ is relatively more attractive than riskless Asset~$i$, then the right hand side above is one at most, and we obtain
	$$
	\frac{1+r}{\mathbb{E}[1+X_j(0)]} \le 1
	$$
	from which it follows that
	$$
	\mathbb{E}[X_j(0)] \ge r .
	$$
	Thus, $\mathbb{E}[X_j(0)]\ge r$ is necessary, but not sufficient
	for risky Asset~$j$ to be relatively more attractive than riskless Asset~$i$. $(v)$ The reader should not confuse the definition of dominant asset with the definition of stochastic dominance. Recall that stochastic dominance involves only the marginal distributions of two random variables, while the dominant asset definition involves the correlation between $[1 + X_j (0)]^{ -1}$ and $1 + X_i(0)$, which depends on the joint distribution of $X_j(0)$ and $X_i(0)$.

	\vspace{6mm}
	\section{Dominant Asset Theorem}
	\label{SECTION: DOMINATE ASSET THEOREM}
	\vspace{-1mm}
	The theorem below tells us that the satisfaction of the dominant asset inequality leads to an optimal portfolio which involves investing  100\% of available funds in a single asset. In other words, if an asset  is dominant, ``bet the farm" on it.
	
	
	\vspace{5mm}
	{\bf Dominant Asset Theorem}: {\it Given a collection of $m$ assets, if Asset $j$ is dominant, 
		then, for all $n \geq 1$, $g_n(K)$ is maximized~by 
		$$
		K_n^* = e_j
		$$
		where $e_j$ is the unit vector in the $j$-th coordinate direction. 
		Furthermore, the resulting optimal expected logarithmic growth rate is given by
		\[
		g_n^* =  g_1^* = \mathbb{E}\left[ {\log (1 + {X_j}(0))} \right].
		\]
	}

	\vspace{-1mm}
	{\bf Proof}: In order to prove $K_n^* = e_j$, it suffices to show that~\mbox{$g_n(K) \leq g_n(e_j)$} for  $K \in \cal{K}$. For notational convenience, we work with the random vector
	$$
	\mathcal{R}_n \doteq \mathcal{X}_n + \bf{1}
	$$ representing the total return with~$i$-th
	component $\mathcal{R}_{n,i}$. Letting \mbox{${\bf 1} \doteq [1 \;\; 1\;\; \cdots \; 1]^T \in \mathbb{R}^m$}, since~\mbox{$K^T {\bf 1} = 1$} for~\mbox{$K \in \mathcal{K}$}, it follows that
	\begin{eqnarray*}
		g_n(K)\!\!\! &=& \!\!\! \frac{1}{n} \mathbb{E}[\log(1+K^T 	\mathcal{X}_n)]\\[5pt] 
		&=& \!\!\!\frac{1}{n} \mathbb{E}[\log(K^T \mathcal{R}_n)].
	\end{eqnarray*}
	Hence, by applying Jensen's inequality to the concave logarithm function above, we obtain a chain of inequalities
	\begin{eqnarray*}
		{g_n}(K) - {g_n}({e_j}) \!\!\! &=& \!\!\! \frac{1}{n} \mathbb{E}\left[ {\log \frac{{{K^T}{\mathcal{R}_n}}}{{{\mathcal{R}_{n,j}}}}} \right] \\ [5pt]
		& \leq & \!\!\! \frac{1}{n}\log \mathbb{E}\left[ {\frac{{{K^T}{\mathcal{R}_n}}}{{{\mathcal{R}_{n,j}}}}} \right]\\
		&=& \!\!\! \frac{1}{n} \log \left( \sum\limits_{i = 1}^m {{{K}_i}\mathbb{E}\left[ {\frac{{{\mathcal{R}_{n,i}}}}{{{\mathcal{R}_{n,j}}}}} \right]} \right) \hfill \\[5pt]
		&=&\!\!\!  \frac{1}{n} \log \left( \sum\limits_{i = 1}^m {{{K}_i}\mathbb{E}\left[ {\prod\limits_{k = 0}^{n - 1} {\frac{{1 + {X_i}\left( k \right)}}{{1 + {X_j}\left( k \right)}}} } \right]} \right)  \hfill \\[5pt]
		&=&\!\!\! \frac{1}{n} \log  \left( \sum\limits_{i = 1}^m {{{K}_i}{{\left( {\mathbb{E}\left[ {\frac{{1 + {X_i}\left( 0 \right)}}{{1 + {X_j}\left( 0 \right)}}} \right]} \right)}^n}} \right) \\ [5pt]
		&\leq&\!\!\!  \frac{1}{n} \log 1\\
	\end{eqnarray*} 
	where the last inequality follows from the dominance of Asset $j$ and the fact that $K \in \mathcal{K}$. Now, since $\log 1  = 0$, it follows that
	$$
	g_n(K) \leq g_n(e_j)
	$$
	~and 
	$$
	g_n^*  = g_n(e_j).
	$$
	To complete the proof, it remains to show that~\mbox{$g_n^* = g_1^*$}. This is easily obtained by recalling that the $X_i(k)$ are i.i.d. and observing that
	\begin{eqnarray*}
		g_n^* \!\!\! &=& \! \! \! g_n(e_j)\\[5pt] 
		&=& \! \! \! \frac{1}{n}\mathbb{E}[ \log \mathcal{R}_{n,j} ]
		\\ [5pt]
		& = & \!\!\! \frac{1}{n} \mathbb{E}\biggl[\log \prod_{k=0}^{n-1} (1+X_j(k)) \biggr]
		\\ [5pt]
		& = & \!\!\! \frac{1}{n} \sum_{k=0}^{n-1}  \mathbb{E} \left[ \; \log \left( 1+X_j(k) \right) \; \right]
		\\ [5pt]
		& = & \!\!\! \mathbb{E}\bigl[\log\bigl(1+X_j(0)\bigr)\bigr] \\[5pt]
		&=& \!\!\! g_1(e_j)\\[5pt]
		& =& \!\!\! g_1^*.  \;\;\;\; \square
	\end{eqnarray*}

	\vspace{5mm}
	\section{Application to Stock-Market Data}
	\label{SECTION: EMPIRICAL DATA STUDY}
\begin{figure}[htbp]
		\vspace{-0mm}
		\begin{center}
			\graphicspath{{Figs/}}
			\includegraphics[scale=0.4]{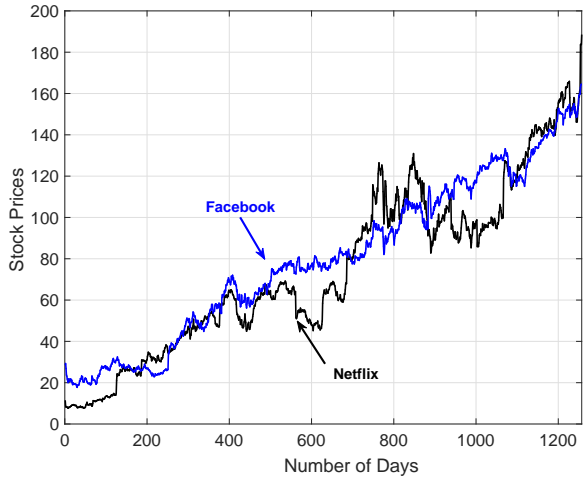}
			\figcaption{Stock Prices of Facebook and Netflix}
			\label{fig:Stock_Prices_FB_NFLEX}
		\end{center}
		\vspace{0mm}
	\end{figure}

	\vspace{-0mm}
	Per discussion in the introduction, in this section, our objective is to illustrate some of the practical issues which arise when studying asset domination using  historical data. To this end, we consider three assets as  portfolio candidates. Asset~1 is  Netflix~\mbox{(ticker: NFLX)}, Asset~2  is Facebook (ticker: FB) and Asset~3 is a  riskless asset with   daily interest rate~$r \geq 0$. 
	We consider the problem of rebalancing our positions in these assets  over a four-year period beginning on January~24,~2013. We work with the adjusted daily closing prices for this period and demonstrate how the Dominant Asset Theorem might apply in practice. The price plot for the  two stocks in Figure~\ref{fig:Stock_Prices_FB_NFLEX} begins with the~126-day period prior to the start of trading. This data was used as a ``training set'' to initialize the analysis to follow.
	
	\vspace{3mm}
	Given the fact that a practitioner should rightfully view the  stochastic process model for the stock returns $X_i(k)$  as nonstationary, when testing for satisfaction of the relative attractiveness inequality, we work with a sliding window  consisting of the most recent~$N$ trading days. Hence, at day~$k$, we use an empirical estimate of the expected value for the attractiveness ratio involving the~$i$-th and~$j$-th assets which is given by
	$$
	R_{ij}(k) \doteq \frac{1}{N}\sum_{\ell = 0}^{N-1}\frac{1 + X_i(k-\ell)}{1 + X_j(k - \ell)}.
	$$
	The simulations to follow use a window size of~$N = 126$ which corresponds to about six months. Beginning with the initial condition~$R_{ij}(0)$ established using the training set, we generate the~$R_{ij}(k)$ over the period of interest. In view of the nonstationarity of the returns,  as seen in the simulation to follow, the $R_{ij}(k)$ are time-varying. Hence, an asset which is dominant at some point in time may no longer be dominant at a later time.
	
	\vspace{4mm} 
	With the considerations above, we begin our analysis with~$r=0 $ and  consider the following two questions. {\it Question~1}:  In a zero interest rate environment, over what time periods is Netflix the dominant asset? During such periods, in accordance with the theorem, the trader is non-diversified with the entire portfolio in Netflix. {\it Question~2}: At stage~$k$, how  large must the interest rate~$r$ be so that the riskless asset is dominant?   That is, when the interest rate is suitably high, our theory dictates that the trader has a portfolio which is 100\% in fixed income with no positions in Netflix and Facebook. 
	
\begin{figure}[htbp]
		\vspace{-0mm}
		\begin{center}
			\graphicspath{{Figs/}}
			\includegraphics[scale=0.40]{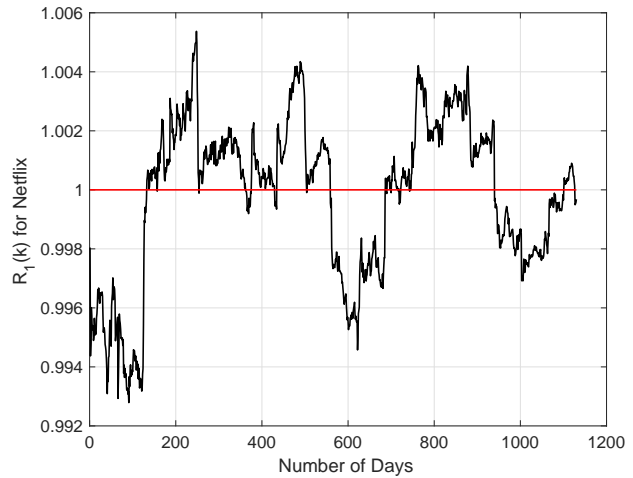}
			\figcaption{The Relative Attractiveness Plot for Netflix}
			\label{fig:Relative_Attractiveness_Plot_for_Netflix}
		\end{center}
		\vspace{-0mm}
	\end{figure}

	\vspace{4mm}
	To answer the first question, we provide a plot of 
	$$
	R_1(k) \doteq \max\{R_{21}(k),R_{31}(k)\}
	$$
	versus $k$ in Figure~\ref{fig:Relative_Attractiveness_Plot_for_Netflix},
	and, consistent with the theorem, we deem Netflix to be~{\it dominant} over the subset of time periods for which
	$
	R_1(k) \leq 1.
	$
	Over the periods when~\mbox{$R_1(k) > 1$}, there are various additional scenarios which can be studied with the given data. For example, sometimes there is no dominant asset and at other times either Facebook or the riskless asset is dominant. 
	
\begin{figure}[htbp]
		\vspace{-0mm}
		\begin{center}
			\graphicspath{{Figs/}}
			\includegraphics[scale=0.4]{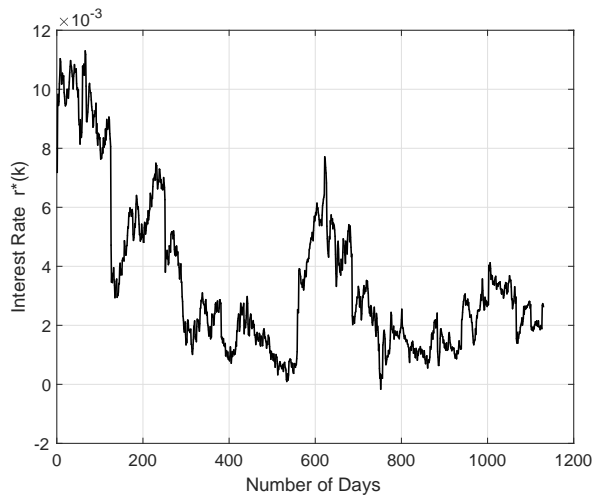}
			\figcaption{Interest Rate for Dominance of Riskless Asset}
			\label{fig:Relative_Attractiveness_Plot_for_Riskless_Asset}
		\end{center}
		\vspace{-2mm}
	\end{figure}

	\vspace{4mm}
	To answer the second question, we begin with the following observation: The theorem tells us that Asset 3, the riskless asset, is dominant if and only if
	$$
	\max\{\mathbb{E}[X_1(0)],\mathbb{E}[X_2(0)]\} \leq r.
	$$ 
	Again taking the nonstationarity of the data into account, we let~$r^*(k)$ denote the estimated value of the left hand size above based on the most recent $N$-day window. That is, we~take
	$$
	r^*(k) \doteq \frac{1}{N}\max \left\{\sum_{\ell = 0}^{N-1} X_1(k-\ell), \sum_{\ell = 0}^{N-1} X_2(k-\ell)\right \}.
	$$
	In Figure~\ref{fig:Relative_Attractiveness_Plot_for_Riskless_Asset},
	we see that there are time periods when the market is performing quite well and it takes a remarkably high interest rate in order to forego investing in the stocks. 
	
	\vspace{6mm}
	\section{Conclusion and Future Work}
	\label{SECTION: CONCLUSION AND FUTURE WORK}
	\vspace{-0mm}
	In this paper, we showed that if an asset is dominant, an optimal trading strategy is to invest all available funds into it. 
	For such cases,   rebalancing becomes moot and the trading performance, namely $g_n^*$, is a constant. It is also worth mentioning that the paradigm presented in this paper remains valid for a wide variety of other choices for the admissible feedback gain set~$\mathcal{K}$ associated with leveraged investments with some components $K_i > 1$. For example, in~\cite{Hsieh_Barmish_Gubner_2016_CDC}, this is achieved by imposing a   ``survival" constraint which disallows any trade that can potentially lead  to~\mbox{$V(k) < 0$.}  
	
	\vspace{4mm}
	Regarding further research, one obvious continuation would be to study the case when $K_i<0$ is allowed. As mentioned in Section~\ref{SECTION: PROBLEM FORMULATION}, this corresponds to short selling. In this situation, we envision a similar definition of dominant asset and results along the lines of those given here.   A second direction for further work involves the case when  no dominant asset exists, Might it ever be true that  $g_n^* > g_1^*$?  We believe that an affirmative answer to this question would be important. It would tell us that
	a low-frequency trader such as a buy and holder might  strictly outperform the high-frequency trader. 
	
	\vspace{4mm}
	Finally, it would be important to develop  new results on Kelly-based trading which do not rely knowledge of a perfect stochastic model for the returns $X_i(k)$.
	For cases when the model is either partially known or  completely unknown, we plan to investigate the extent to which the theory in this paper can be extended. Our preliminary work along these lines suggests that there may be a more general version of the Dominant Asset Theorem which is established using asymptotic analysis to obtain performance guarantees for~$n$  suitably large.
	


\end{document}